 \documentclass[preprint2]{aastex}

\newcommand{\jan}{Jan10}
\newcommand{\sep}{Sep09}

\newcommand{\six}{HD~80606}
\newcommand{\seven}{HD~80607}
\newcommand{\sixb}{HD~80606b}
\newcommand{\lc}{light curve}
\newcommand{\lcs}{light curves}

\newcommand{\Lcs}{Light curves}

\newcommand{\oot}{out-of-transit}

\newcommand{\spitzer}{{\it Spitzer}}

\newcommand{\fov}[2]{\ensuremath{ \rm #1.\!^\prime #2 \times \rm #1.\!^\prime #2}}
\newcommand{\fova}[4]{\ensuremath{ \rm #1.\!^\prime #2 \times \rm #3.\!^\prime #4}}
\newcommand{\pxs}[2]{\ensuremath{#1.\!^{\prime\prime}#2\ \rm{pixel}^{-1}}}
\newcommand{\bin}[1]{\ensuremath{ \rm #1 \times \rm #1} \rm{pixel}}

\newcommand{\figr}[1]{Fig.~\ref{fig:#1}}
\newcommand{\secr}[1]{\mbox{\S\ \ref{sec:#1}}}
\newcommand{\eqr}[1]{Eq.~\ref{eq:#1}}

\newcommand{\tabr}[1]{\mbox{Table~\ref{tab:#1}}}

\slugcomment{Accepted for publication in ApJ}

\shorttitle{Observations of two transits of HD~80606b}

\shortauthors{Shporer et al.}

\begin{document}

\title{Ground-based multisite observations of two transits of HD~80606b}

\author{A.~Shporer\altaffilmark{1, 2}, 
J.~N.~Winn\altaffilmark{3}, 
S.~Dreizler\altaffilmark{4}, 
K.~D.~Col\'{o}n\altaffilmark{5}, 
W.~M.~Wood-Vasey\altaffilmark{6}, 
P.~I.~Choi\altaffilmark{7}, 
C.~Morley\altaffilmark{8}, 
C.~Moutou\altaffilmark{9}, 
W.~F.~Welsh\altaffilmark{10}, 
D.~Pollaco\altaffilmark{11}, 
D.~Starkey\altaffilmark{12}, 
E.~Adams\altaffilmark{8}, 
S.~C.~C.~Barros\altaffilmark{11}, 
F.~Bouchy\altaffilmark{13,14}, 
A.~Cabrera-Lavers\altaffilmark{15,16}, 
S.~Cerutti\altaffilmark{6}, 
L.~Coban\altaffilmark{6}, 
K.~Costello\altaffilmark{6}, 
H.~Deeg\altaffilmark{15,17}, 
R.~F.~D\'{\i}az\altaffilmark{13}, 
G.~A.~Esquerdo\altaffilmark{18}, 
J.~Fernandez\altaffilmark{4}, 
S.~W.~Fleming\altaffilmark{5}, 
E.~B.~Ford\altaffilmark{5}, 
B.~J.~Fulton\altaffilmark{1}, 
M.~Good\altaffilmark{6}, 
G.~H\'ebrard\altaffilmark{13}, 
M.~J.~Holman\altaffilmark{18}, 
M.~Hunt\altaffilmark{6}, 
S.~Kadakia\altaffilmark{10}, 
G.~Lander\altaffilmark{6}, 
M.~Lockhart\altaffilmark{8}, 
T.~Mazeh\altaffilmark{19}, 
R.~C.~Morehead\altaffilmark{5}, 
B.~E.~Nelson\altaffilmark{5}, 
L.~Nortmann\altaffilmark{4}, 
F.~Reyes\altaffilmark{5}, 
E.~Roebuck\altaffilmark{6}, 
A.~R.~Rudy\altaffilmark{7}, 
R.~Ruth\altaffilmark{5}, 
E.~Simpson\altaffilmark{11}, 
C.~Vincent\altaffilmark{6}, 
G.~Weaver\altaffilmark{6}, 
J.-W.~Xie\altaffilmark{5}} 
\altaffiltext{1}{Las Cumbres Observatory Global Telescope Network, 6740 Cortona Drive, Suite 102, Santa Barbara, CA 93117, USA; ashporer@lcogt.net}
\altaffiltext{2}{Department of Physics, Broida Hall, University of California, Santa Barbara, CA 93106, USA}
\altaffiltext{3}{Department of Physics and Kavli Institute for Astrophysics and Space Research, Massachusetts Institute of Technology, Cambridge, MA 02139, USA}
\altaffiltext{4}{Georg-August-Universit\"at, Institut f\"ur Astrophysik, Friedrich-Hund-Platz 1, 37077 G\"ottingen, Germany}
\altaffiltext{5}{Department of Astronomy, University of Florida, 211 Bryant Space
Science Center, Gainesville, FL 32611-2055, USA}
\altaffiltext{6}{University of Pittsburgh, Department of Physics and Astronomy, 3941 O'Hara St., Pittsburgh, PA 15260, USA}
\altaffiltext{7}{Department of Physics and Astronomy, Pomona College, 610 N College Ave, Claremont, CA 91711, USA}
\altaffiltext{8}{Department of Earth, Atmospheric, and Planetary Sciences, Massachusetts Institute of Technology, 77 Massachusetts Avenue, Cambridge, MA 02139, USA}
\altaffiltext{9}{Laboratoire d'Astrophysique de Marseille, Universit\'e de Provence, CNRS (UMR 6110), 38 rue Fr\'ed\'eric Joliot Curie, 13388 Marseille cedex 13, France}
\altaffiltext{10}{Department of Astronomy, San Diego State University, 5500 Campanile Drive, San Diego, CA 92182, USA}
\altaffiltext{11}{Astrophysics Research Centre, School of Mathematics \& Physics, QueenÕs University, University Road, Belfast, BT7 1NN, UK}
\altaffiltext{12}{DeKalb Observatory H63, Auburn, IN 46706, USA}
\altaffiltext{13}{Institut d'Astrophysique de Paris, UMR7095 CNRS, Universit\'e Pierre \& Marie Curie, 98bis boulevard Arago, 75014 Paris, France}
\altaffiltext{14}{Observatoire de Haute-Provence, CNRS/OAMP, 04870 Saint-Michel-l'Observatoire, France}
\altaffiltext{15}{Instituto de Astrof\'\i sica de Canarias, C. Via Lactea S/N, 38205 La
Laguna, Tenerife, Spain}
\altaffiltext{16}{GTC Project Office, E-38205 La Laguna, Tenerife, Spain}
\altaffiltext{17}{Universidad de La Laguna, Dept. de Astrof\'\i sica,  38200 La Laguna, Tenerife, Spain}
\altaffiltext{18}{Harvard-Smithsonian Center for Astrophysics, Cambridge, MA 02138, USA}
\altaffiltext{19}{Wise Observatory, Tel Aviv University, Tel Aviv 69978, Israel}

\begin{abstract}

  We present ground-based optical observations of the September 2009
  and January 2010 transits of \sixb.  Based on 3 partial \lcs\ of the
  September~2009 event, we derive a midtransit time of
  $T_c$~[HJD] = 2455099.196 $\pm$ 0.026, which is about
  1$\sigma$ away from the previously predicted time.  We observed the
  January 2010 event from 9 different locations, with most phases of
  the transit being observed by at least 3 different teams. We
  determine a midtransit time of $T_c$~[HJD] = 2455210.6502 $\pm$
  0.0064, which is within 1.3$\sigma$ of the time derived from a
  \spitzer\ observation of the same event.

\end{abstract}

\keywords{planetary systems --- stars: individual (HD~80606)}

\section{Introduction}

The sample of transiting exoplanets has grown rapidly in recent years,
but \sixb\ stands out from this crowd by virtue of its long period
(111.4~days) and high orbital eccentricity ($e$=0.93). Initially the
planet was discovered in a Doppler survey and was not known to transit
(Naef et al.~2001). However, the planet is near pericenter during
superior conjunctions, increasing the probability of occultations
(secondary eclipses). This motivated the observations of
\cite{laughlin09}, who found that the orbital inclination is indeed
close enough to 90$^\circ$ for occultations to occur.  Soon after, the
discovery of a transit (primary eclipse) was reported by several
authors (\citealt{moutou09, fossey09, garcia09}). In addition, the
orbit was shown to be misaligned with the plane defined by the stellar
rotation \citep{moutou09, pont09, winn09, hebrard10}.  This system has
become an important case study for theories of orbital migration, tidal
interactions, and giant planet atmospheres \citep[e.g.,][]{wu03,
  fabrycky07, triaud10,pont08,laughlin09,knutson09}.

For any transiting planet, it is desirable to observe multiple
transits. This allows the system parameters to be refined, especially
the orbital period. In addition, a deviation from strict periodicity
may be the signature of additional bodies in the system, such as
another planet or a satellite \citep[e.g.,][]{holman05, agol05,
  simon07, nesvorny10}.

In the case of \sixb\, observations of complete transits are
challenging. This is partly because transits are rare, occurring once
every 111.4 days, and also because the transit duration is nearly 12
hours, making it impossible to be observed in entirety from a single
ground-based site. \cite{hidas10} and \cite{winn09} carried out
multisite campaigns to achieve more complete coverage of the
transit. This paper reports the results of our more recent
observations. In September~2009, only the first portion of the transit
could be observed, but our combined light curve for January~2010
ranges over the entire transit, with most phases having been observed
with at least 3 different telescopes.

\cite{hebrard10} utilized the \spitzer\ observatory to obtain a high
quality \lc\ of the entire January~2010 transit event. Besides deriving
refined system parameters, they measured a midtransit time earlier by
3$\sigma$ than was predicted by \cite{winn09}. This discrepancy
suggested the intriguing possibility of the existence of a third body
in the \six\ transiting system. The primary goal of our analysis was
to determine the times of these events as accurately as
possible. Although our results are not as precise as the \spitzer\
result, we provide an independent estimate of the midtransit time. Our
observations and photometric processing are described in
\secr{obs}. In \secr{dataanal} we describe our analysis, the results
of which are presented in \secr{res} and discussed in \secr{dis}.

\section{Observations}
\label{sec:obs}

The new data presented in this paper are 3 photometric time series
spanning the first portion of the transit of 2009~September 23/24 (hereafter \sep),
based on observations across North America; and 10 photometric time
series spanning the entire transit of 2010~January~13/14  (hereafter \jan), based on
data from Israel, Europe, Canary Islands, North America and
Hawaii.

In all cases we gathered CCD images encompassing the target star,
\six\ (G5V, $V$=9.06, $B-V$=0.76), and its visual binary companion,
\seven\ (G5V, $V$=9.07, $B-V$=0.87) which is located 21\arcsec\ east.
The similarity in brightness and color between the two stars
facilitates differential photometry. Since the stars are relatively
bright, we defocused most of the telescopes that were used, thereby
allowing for longer exposures without saturation, and reducing the
impact of pixel sensitivity variations and seeing variations. We
always ensured that the PSFs of the two stars were well separated.
Brief descriptions of the observations from each site are given below,
along with an abbreviated observatory name that we will use in the
remainder of this paper.

{\it Wise Observatory\footnote{ \url{http://wise-obs.tau.ac.il/}}
  (WO), Israel.} The target was observed with the 1.0~m telescope for
9~hours on the \jan\ transit night, completely covering ingress, and
for 2~hours on the following night. An RGO $Z$ filter was used and the
telescope was slightly defocused. The detector was a back-illuminated
Princeton Instruments CCD, with a \fova{12}{6}{13}{0} FOV and a pixel scale of
\pxs{0}{580}.

{\it Gran Telescopio Canarias\footnote{
    \url{http://www.gtc.iac.es/en/pages/gtc.php}} (GTC), La Palma,
  Canary Islands.}  We observed the target with the Optical System for
Imaging and low Resolution Integrated Spectroscopy (OSIRIS), mounted
on the 10.4~m GTC telescope. The observations lasted for 8.3~hours on the \jan\ transit night,
all within the transit. A small defocus was applied, resulting in a
typical PSF FWHM of 1.5\arcsec. We used the narrow band imager with
the red range tunable filter, alternating between four narrow bands in
the range 7680--7780~\AA, each with a FWHM of 12~\AA. The instrument
has a maximum FOV of \fov{7}{8} and a pixel scale of
\pxs{0}{127}. Analysis of the \lc\ for each individual narrow band
filter are presented in \cite{colon10}; here, we are concerned with the
\lc\ based on a summation of the flux observed in all four filters. We
adopt the SDSS-$i'$ band limb darkening coefficients when fitting this
\lc.

{\it Observatoire de Haute Provence\footnote{
    \url{http://www.obs-hp.fr}} (OHP), France.}  The target was
observed with the 1.2~m telescope for 4.3~hours during the first half
of the \jan\ transit, with no \oot\ observations.  As was the case with the 
spectroscopic observations secured simultaneously with Sophie at OHP
\citep{hebrard10}, the transit sequence had to be stopped due to
cloudy weather. Photometric observations were also secured on several nights
shortly before and after the transit night, to put constraints on the stellar
activity. These data are described in more details by
\cite{hebrard10}. No defocus was applied here; instead, to allow for
longer exposures, we used a neutral density filter along with a
Gunn-$r$ filter.

{\it Allegheny Observatory\footnote{
    \url{http://www.pitt.edu/$\sim$aobsvtry/}} (AO), Pittsburgh,
  Pennsylvania, USA.}  We used the 0.41~m (16~in) Meade telescope and
a SBIG \pxs{0}{56} CCD, with a FOV of
\fova{20}{0}{30}{0}. Observations of the \jan\ transit were done in
the $I$ filter with no defocus, from the time of second contact until
about 2~hours after the transit ended, a total of 11.4~hours.

{\it Rosemary Hill Observatory\footnote{
    \url{http://www.astro.ufl.edu/information/rho.html}} (RHO),
  Bronson, Florida, USA.}  We used the 0.76~m Tinsley telescope and a
SBIG ST-402ME CCD camera mounted at the f/4 Newtonian focus, with a
FOV of \fova{3}{88}{2}{56} (half of the entire FOV). During the \jan\
event we observed for 4.9~hours with no \oot\ data. We used the same
instrument to observe the \sep\ event, obtaining 2.4~hours of data
just before the transit started, which helps constrain the transit
start time. Both events were observed with the SDSS-$i'$ filter and
the telescope was defocused.

{\it Fred Lawrence Whipple Observatory\footnote{
    \url{http://www.sao.arizona.edu/FLWO/whipple.html}} (FLWO),
  Mt.~Hopkins, Arizona, USA.}  We used Keplercam, mounted on the 1.2~m
telescope, and observed the target for 2.8~hours during the flat
bottom part of the \jan\ transit. The same instrument was also used to
observe the very beginning of the \sep\ event, obtaining 1.8~hours of
data. KeplerCam includes a 4K $\times$ 4K FairChild Imaging CCD486,
with a pixel scale of \pxs{0}{672} (2 $\times$ 2 binning) and a
\fov{23}{1} FOV. FLWO observations of both transit events used the
SDSS-$i'$ filter and started immediately after the target rose.

{\it Table Mountain Observatory\footnote{
    \url{http://tmoa.jpl.nasa.gov/}} (TMO), Wrightwood, California,
  USA.}  We used the 1.0~m telescope with an Apogee U16 4K $\times$ 4K
CCD, with a pixel scale of \pxs{0}{18} and a \fov{12}{3} FOV. The
SDSS-$i'$ filter was used here. TMO observations completely cover the
\jan\ egress, with a total of 6.6~hours, of which 2.9~hours are after
fourth contact.

{\it George R.~Wallace Jr.~Astrophysical Observatory\footnote{
    \url{http://web.mit.edu/wallace/}} (WAO), Westford, Massachusetts,
  USA.} We used two identical 0.36~m telescopes on the \jan\ transit
night, each equipped with a SBIG STL-1001E 1K $\times$ 1K CCD camera
with a pixel scale of \pxs{1}{29}, resulting in a FOV of
\fov{21}{5}. We refer to these instruments as WAO-1 and WAO-2. At both
telescopes we observed in the I filter with no defocus. With WAO-1 we
were able to observe for 15~minutes during the flat bottom part,
before observing conditions degraded. Observations resumed with both
instruments at the time of third contact and continued for 4.2~hours,
of which 2.4~hours were \oot.

{\it Faulkes Telescope North\footnote{
    \url{http://faulkes-telescope.com/}} (FTN), Mt.~Haleakala, Maui,
  Hawaii, USA.}  We used the LCOGT\footnote{\url{http://lcogt.net}}
2.0~m FTN telescope, and the Spectral Instruments camera with the
Pan-STARRS $Z$ filter. The camera has a back-illuminated Fairchild
Imaging CCD and we used the default \bin{2} binning mode, with an
effective pixel scale of \pxs{0}{304}. The telescope was defocused and
the \fov{10}{5} FOV was positioned and rotated so the guiding camera
FOV will contain a suitable guide star. We observed the target on the
\jan\ transit night for 9.2~hours, from the beginning of egress until
7~hours after the transit ended. We observed also during the two
adjacent nights, for 2~hours on the preceding night (Jan.~12/13), and
for 3.5~hours on the following night (Jan.~14/15).

{\it Mount Laguna Observatory\footnote{
    \url{http://mintaka.sdsu.edu/}} (MLO), San Diego, CA.}  Only the
\sep\ event was observed. We used the 1.0~m telescope with a Fairchild
Imaging 2K $\times$ 2K CCD and SDSS-$i'$ filter. Observations were
done with a 300 $\times$ 300 pixel sub-array and a FOV of
\fov{2}{0}. The telescope was defocused and the very beginning of the
event was observed, for 1.6~hours with no \oot\ data.

Four additional \lcs\ were obtained for the \jan\ event by the
Liverpool Telescope, MONET-North telescope, DeKalb Observatory, and a
third telescope at WAO, and one for the \sep\ event by AO. However,
those data displayed a very high noise level and strong systematic
effects, and were not included in our subsequent analysis.

The CCD data were reduced using standard routines for bias
subtraction, dark current subtraction (when necessary) and flat-field
correction. We used aperture photometry to derive the flux of HD~80606
and HD~80607, and divided the former by the latter to obtain a time
history of the flux ratio, which we refer to as the light curve. We
took care to choose aperture sizes to avoid contamination of one
stellar signal by the other star. Our time stamps represent the
Heliocentric Julian Date, based on the UTC at midexposure (and not the
uniformly flowing terrestrial time system advocated by
\citet{eastman10}).  This is also the time system that was used for
the \spitzer\ analysis of \citet{hebrard10}.

All light curves were averaged into 10~minute bins, using 3$\sigma$
outlier rejection. The error bar assigned to each data point was 
the standard deviation of the mean of all the measurements
contributing to each 10~minute bin, which ranged in number from 4 to
63.  There is no significant information loss due to time binning,
because the binning time of 10~minutes is shorter than the duration of
ingress and egress by more than an order of magnitude.

The data cover only the first portion of the 2009~September~23/24
event, but they provide complete coverage of the 2010~January~13/14
event.  Although the first hour of the transit was observed from only
one observatory (WO), the rest of the transit was observed by 3--5
different sites, which is very helpful for identifying and decreasing the 
influence of any systematic effects that are specific to each observatory (i.e.\
correlated noise, or red noise) which is frequently a problem with
ground-based photometric data \citep[e.g.,][]{pont06, carter09,
  sybilski10}.

\tabr{lc} gives the photometric data that were obtained and analyzed.
Each data point represents a 10~minute binned average of the flux
ratio of HD~80606 to HD~80607.  The normalization factors and error
rescaling factors that are described in \secr{dataanal} were {\it not}
applied to the data given in the table.  \tabr{obs} gives a list of
all the observatories.

\section{Data Analysis}
\label{sec:dataanal}

In this section we describe the methods by which we combined the data
and derived the midtransit time of each event. We describe the process
in detail for the \jan\ event; the details were very similar for the
\sep\ data.

Because the quality of the \spitzer\ \lc\ is superior to any
ground-based \lc, we did not attempt to use our data to refine the
basic system parameters other than the midtransit times.  Instead, we
used the parameters derived by \cite{hebrard10} as constraints on the
light curve shape, while allowing the midtransit time to be a free
parameter, as described below.

A simultaneous analysis of several \lcs\ obtained by different
instruments is a challenging task. \cite{winn09} and \cite{hidas10}
have carried out a similar task, although for a smaller number of data
sets. One of the crucial points is the placement of the different flux
ratio \lcs\ onto the same scale, despite the differences in bandpasses,
detectors, and weather conditions at each observatory. One way of
thinking about this problem is that we need to establish the
out-of-transit flux ratio that was measured, or that would be
measured, by each observatory. \citet{winn09} performed this
calibration by using data taken on nights when the planet was not
transiting, a method that may be affected by night-to-night variations
due to varying observing conditions. \cite{hidas10} allowed the
overall flux ratio scale to be a free parameter for each \lc, thereby
increasing the overall number of fitted parameters. Here we chose an
intermediate approach. We assigned a normalization factor to each
\lc, estimating it from the \oot\ data whenever possible, and
allowing it to be a free parameter when there was insufficient \oot\
data.

We divided our \lcs\ into three groups, based on the amount of \oot\
data:

\emph{Group I} includes the light curves with abundant \oot\
information. The only members of this group are the WO and FTN \lcs\
of the \jan\ event, where \oot\ measurements were obtained on the
transit night and on one (for WO) or two (for FTN) of the adjacent
nights. For each of the two \lcs\ separately, we subsequently fitted a
2nd degree polynomial to the \oot\ flux ratio points vs.~airmass, time
and PSF FWHM, and divided the entire \lc\ by this polynomial. The
effect of this process on the in-transit points was small, typically
at the few 0.01\% level. The assigned normalization factor for each of
the two resulting \lcs\ was 1.0.

\emph{Group II} includes \lcs\ that have at least 1.5~hours of \oot\
measurements. Their normalization factors were taken to be the mean
\oot\ flux ratio. This group includes 4 \lcs: the AO, TMO, and the two
WAO \lcs.

\emph{Group III} includes the remaining 4 \lcs\ (GTC, OHP, RHO and
FLWO) with either a small amount of data or no \oot\ data at all. The
normalization factors were taken to be free parameters in our model.

Our model for the data is based on the premise of two spherical
objects, a non-luminous planet and a limb-darkened star with a
quadratic limb darkening law, in an eccentric Keplerian orbit.  For
each binned time stamp we calculated the sky projected planet-star
distance and used the equations of \cite{mandel02} to calculate the
relative flux at that time. Our code accounts for the light travel
time effects described by \cite{hebrard10}.

The model included a total of 34 parameters: orbital period $P$,
planet-to-star radius ratio $r=R_p/R_s$, orbital semimajor axis in
units of the stellar radius $a/R_s$, orbital eccentricity $e$,
argument of periastron $\omega$ (in fact our fitting parameters were
actually $e\cos\omega$ and $e\sin\omega$), inclination angle $i$, an
individual periastron passage time $\{T_{p,i}\}^{10}_{i=1}$ for each
of the 10 \lcs\ (which were later converted into midtransit times), two
limb darkening coefficients $u_1$ and $u_2$, for each of the different
4 filters used, and 10 normalization factors, one for each \lc.

The limb darkening coefficients were estimated from the grids of
\cite{claret00, claret04} for a star with T$_{\rm eff}$=5645~K, $\log
g$=4.5 and [Fe/H]=0.43 \citep{naef01}, and were held fixed in the
fitting process as the data are insensitive to the coefficients. As
mentioned earlier, we used the parameters of \cite{hebrard10} to
constrain the \lc\ shape. We used their values and uncertainties for
$P$, $r$, $a/R_s$, $e\cos\omega$, $e\sin\omega$ and $i$ as a priori
Gaussian constraints by adding penalty terms to the $\chi^2$ fitting
statistics. We constrained in a similar way the normalization factors
of the 6 \lcs\ in Groups I and II described above, while assuming a
normalization factor uncertainty of 0.1\%. This uncertainty is larger
than that of the \lcs\ mean \oot\ flux ratio, typically a few times
0.01\%, and it was used in order to prevent the fitting process from
being dominated by a single \lc.

Out of the 34 model parameters, 8 were held fixed (the limb darkening
coefficients), 12 were controlled mainly by Gaussian priors ($P$, $r$,
$a/R_s$, $e\cos\omega$, $e\sin\omega$, $i$ and 6 normalization
factors), and the remaining 14 were free parameters with uniform
priors ($\{T_{p,i}\}^{10}_{i=1}$ and 4 normalization factors). Our
fitting statistic was:
\begin{equation}
\label{eq:chi2}
\chi^2 = \chi^2_f + \chi^2_{orb} + \chi^2_{norm},
\end{equation}
where the first term on the right hand side is the usual $\chi^2$ statistics:
\begin{equation}
\label{eq:chi2f}
\chi^2_f = \sum_{i=1}^{431} \left[ \frac{f_i(obs) - f_i(model)}{\sigma_{f_i}} \right]^2,
\end{equation}
the second term includes the penalties for the \lc\ parameters (values and uncertainties taken from \citealt{hebrard10}, $P$ in days and $i$ in degrees):
\begin{eqnarray}
\label{eq:chi2orb}
\chi^2_{orb}  = \left[\frac{P - 111.4367}{0.0004}\right]^2 +
\left[\frac{r - 0.1001}{0.0006}\right]^2 +  \nonumber \\
\left[\frac{a/R_s - 97.0}{1.6}\right]^2 +
\left[\frac{e\cos\omega - 0.4774}{0.0018}\right]^2 +  \nonumber \\
\left[\frac{e\sin\omega -  (-0.8016)}{0.0017}\right]^2 + 
\left[\frac{i - 89.269}{0.018}\right]^2,
\end{eqnarray}
and the third term constrained the normalization factors of the 6 \lcs\ for which we have sufficient \oot\ data:
\begin{equation}
\label{eq:chi2norm}
\chi^2_{norm} =  \sum_{l=1}^{6} \left[ \frac{f_{oot,l} - \bar{f}_{oot,l}}{0.001} \right]^2.
\end{equation}

After a preliminary fit, the residuals of each data set were carefully
examined. Two data points were clear outliers, departing from the
model by $>$4$\sigma$, and were rejected, leaving a total of 431
points.  The rejected points were either the first or last data points
in the time series, and were probably affected by the high airmass or
relatively bright twilight sky.  After refitting there were no
additional $>$4$\sigma$ outliers.

Next, for the purpose of determining parameter uncertainties, we
determined appropriate weights for each data point. This was done in
two steps specific to each \lc. First, we rescaled the error bars such
that the median error bar was equal to the rms residual.  We named
this rescaling factor $\alpha$.  If the median error bar was already
equal to or smaller than the rms residual, we set $\alpha=1.0$.  Second,
we attempted to account for correlated (red) noise in each time series
on the critical time scale of the ingress/egress duration (2.8~hours).
We used the ``time-averaging" method \citep[e.g.,][]{winn08}, in which
the residuals are binned using several bin sizes, close to the
duration of ingress and egress. The amount of correlated noise is then
quantified by the ratio between the binned residual \lcs\ standard
deviation and the expected standard deviation assuming pure white
noise. For each \lc\ we took $\beta$ to be the largest ratio among the
bin sizes used, and multiplied the individual error bars by that
factor. We took $\beta$ to be 1.0 when the time-averaging method gave
a smaller value. The values of $\alpha$ and $\beta$ for each \lc\ are
listed on \tabr{obs}.

To determine the ``best'' values of the parameters, and their
uncertainties, we used a Monte Carlo Markov Chain (MCMC) algorithm
\citep[e.g.,][]{tegmark04, ford05} with Metropolis-Hastings
sampling, which has become the standard practice in the literature on
transit photometry \citep{holman06, collier07, burke07}.  We used here
an adaptive approach, similar to the one described by
\cite{shporer09}. The algorithm steps from a multidimensional point in
parameter space, $\bar{P}_i$, to the next, $\bar{P}_{i+1}$, according
to
\begin{equation}
\label{eq:jump}
\bar{P}_{i+1} = \bar{P}_i + f \bar{\sigma} \bar{G}(0,1),
\end{equation}
where $\bar{G}(0,1)$ is a vector of numbers picked randomly from a
Gaussian distribution of zero mean and unit standard deviation,
$\bar{\sigma}$ is a vector of the so-called step sizes, and $f$ (the
only scalar in \eqr{jump}) is a factor chosen to control the fraction
of accepted steps. The value of $f$ was readjusted every $10^3$ steps,
to keep the acceptance fraction near 25\% \citep{gregory05}.  Our
final MCMC included 10 long chains of 500,000 steps each, starting
from different initial positions in parameter space spaced apart by
$\approx$5$\sigma$ from the best-fitting parameters. The posterior
probability distribution of each parameter was constructed from all
long chains after discarding the first 20\% of the steps. We took the
distribution median to be the ``best'' value and the values at the
84.13 and 15.87 percentiles to be the +1$\sigma$ and -1$\sigma$
confidence uncertainties, respectively.

\section{Results}
\label{sec:res}

\begin{figure}
\begin{center}
\includegraphics[scale=0.70]{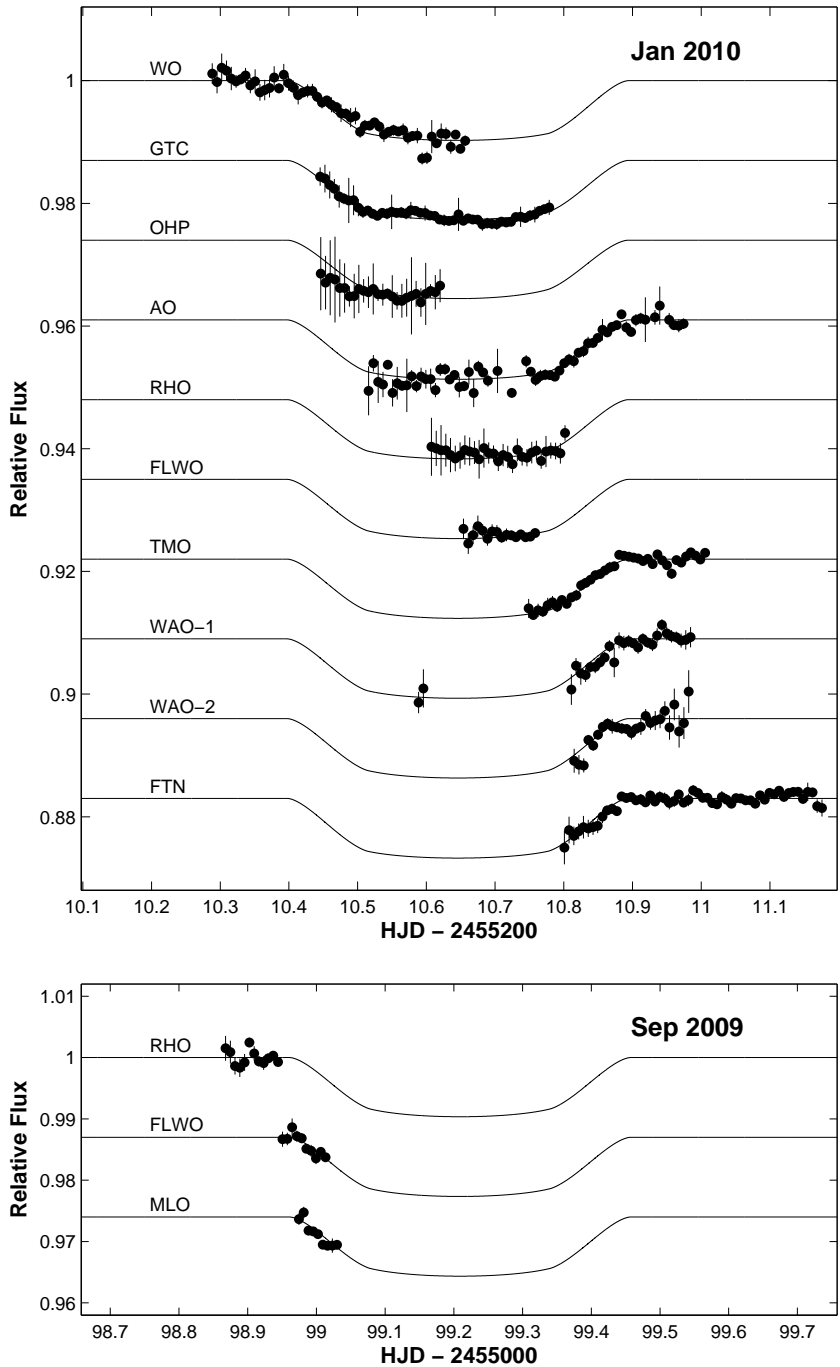}
\caption{\label{fig:lcall1} New light curves of HD~80606.  {\it
    Top.}---The 10 light curves of the \jan\ event.  {\it
    Bottom.}---The 3 light curves of the \sep\ event. Each \lc\ is
  overplotted by the best-fitting model, with limb-darkening
  coefficients appropriate for the observing bandpass.}
\end{center}
\end{figure}

Results of the fitting process are presented visually in \figr{lcall1}
where each \lc\ is plotted separately and overplotted by the fitted
model, using limb darkening coefficients of the corresponding
filter. \figr{lcall2} shows the combined \lc, with the 13 \jan\ and
\sep\ \lcs\ plotted on top of each other.

The fitted model for the \jan\ event has $\chi^2/N_{\rm dof} =442/429$, and
for the \sep\ event $\chi^2/N_{\rm dof}=36/33$. The rms residual of each
binned \lc\ is given in the second-to-last column of \tabr{obs}. The
rightmost column of \tabr{obs} lists the photometric noise rate (PNR)
of the unbinned \lc, defined as the rms divided by $\sqrt{\Gamma}$,
where $\Gamma$ is the median number of data points per unit time
(the ``cadence'').  The PNR is meant to be a quantitative comparison
between the statistical power of different data sets (see also
\citealt{burke08} and \citealt{shporer09}\footnote{We note that the
  equation for the PNR given by \cite{shporer09} includes a typographical 
  error, although their calculations are correct}).

The fitting process resulted in 10 estimates of the periastron passage
time, $\{T_{p,i}\}^{10}_{i=1}$. Using the resulting distributions of
the \lc\ parameters we numerically translated each periastron time to
midtransit time. We then averaged the midtransit times to get our
final estimate of the \jan\ midtransit time, $T_{c,Jan10}$. While
examining the 10 individual midtransit time estimates we noticed that
3 of them have relatively large uncertainties, larger than 0.015 days,
while the rest have uncertainties in the range of 0.005--0.010
days. Therefore we removed them from the sample before
averaging. Those 3 are the OHP, RHO and FLWO \lcs, spanning less than
5~hours without any \oot\ data, hence the increased uncertainties are
expected. We used an unweighted average to get our final midtransit
time estimate, of $T_{c,Jan10}$ [HJD] = 2455210.6502, with an rms of
0.0064 days, close to the typical uncertainty of the 7 individual
estimates. Using regular averaging acts to average out possible
correlated noise affecting individual estimates. Using median or
weighted average changes the midtransit time by $\lesssim
0.25\sigma$. Including the 3 large uncertainty $T_{c,i}$'s mentioned
above results also in a small change to $T_{c,Jan10}$, of less than
0.3$\sigma$ but the scatter is increased by 50\%.

To check the sensitivity of our result to our treatment of correlated
noise, we reran our analysis for the \jan\ data while using
$\beta=1.0$ for all \lcs. The resulting $T_{c,i}$ uncertainties were
smaller by typically 25\%--30\%, and the average midtransit time was
2455210.6476 $\pm$ 0.0093, earlier by 0.4$\sigma$ than our preferred
analysis.

For estimating the midtransit time of the \sep\ event, $T_{c,Sep09}$,
we used only 2 of the 3 $T_{c,i}$'s as the RHO data is entirely
\oot. The average between these two is $T_{c,Sep09} [HJD] =
2455099.196$ and we assign it an uncertainty of 0.026 days, similar to
that of the two $T_{c,i}$'s.

We note that figs.~\ref{fig:lcall1} and \ref{fig:lcall2} were produced
using the average midtransit times of each event, derived above, {\it
  not} with the $T_{c,i}$ of each partial \lc.

\section{Discussion}
\label{sec:dis}

\tabr{tc} lists the midtransit times derived here, the one measured by
\cite{hebrard10} and those predicted by \cite{winn09}. For the \jan\
event \cite{hebrard10} measured a $T_{c,Jan10}$ which is 3$\sigma$
earlier than predicted by \cite{winn09}. Our measured $T_{c,Jan10}$ is
intermediate between those of \cite{hebrard10} and \cite{winn09}. It
is 1.3$\sigma$ later than the \cite{hebrard10} time, and earlier by
1.1$\sigma$ than the \cite{winn09} prediction. Therefore, it does not 
confirm nor refute the earlier transit time measured by \cite{hebrard10}.

There is some discrepancy between \cite{hebrard10} and \cite{winn09}
also in the three parameters determining the \lc\ shape, namely $r$,
$a/R_s$ and $i$. Specifically, the values for the planet-star radius
ratio differ by about 2.5$\sigma$, and for $a/R_s$ and $i$ the
difference is at the 1.5$\sigma$ level. To check which set of values
is preferred by our data we reran our analysis using the \cite{winn09}
values for these three parameters as prior constraints in
\eqr{chi2orb}. This resulted in an increased $\chi^2/N_{\rm dof}$ of 40\%,
showing that our data prefers the \cite{hebrard10} values. This
indicates the discrepancy is not the result of a wavelength dependent
radius ratio and more likely results from underestimated
uncertainties.

\cite{hebrard10} noticed a ``bump'' in their \spitzer\ \lc: a
$\sim$0.1\% increase in flux lasting about 1~hour, just before
midtransit. This bump could be caused by the passage of the planet in
front of a dark spot on the star, although \six\ is not known to be an
active star \citep{hebrard10}. The decreased surface brightness of a
spotted surface element results from decreased temperature relative to
a non-spotted surface element, so the flux increase during a spot
crossing is expected to be wavelength dependent, increasing with
shorter wavelengths \citep{pont08, rabus09}.

\begin{figure}
\begin{center}
\includegraphics[scale=0.5]{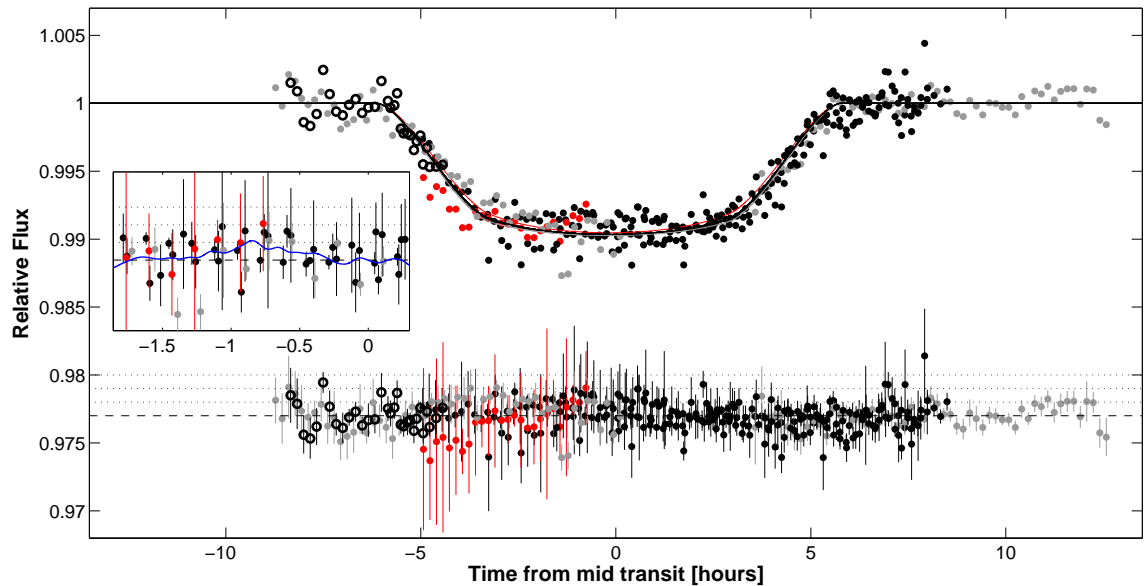}
\caption{\label{fig:lcall2} Composite light curve.  The phase-folded
  light curve based on the 13 new data sets analyzed in this paper.
  \Lcs\ taken with different filters are represented by different
  colors. For each filter the corresponding model \lc\ is plotted as a
  solid line with the same color. Data obtained in the $Z$ band is in
  gray, $i$ and $I$ data in black and $r$ data in red. Open circles
  represent the \sep\ data and filled circles the \jan\ data.  Plotted
  below the light curve are the residuals, with error bars.  The dashed
  line marks the residual zero point, and the 3 dotted lines mark
  relative flux residual levels of 0.1\%, 0.2\% and 0.3\%. The inset
  shows a zoomed-in view of the residuals during the phase when a
  ``rebrightening'' or ``bump'' was observed with \spitzer\ by
  \citet{hebrard10}. The \spitzer\ feature is overplotted in blue. Our
  data neither confirm nor refute the existence of the bump. }
\end{center}
\end{figure}

The inset in \figr{lcall2} shows that our data are not sensitive
enough to identify a 0.1\% flux variation, although it does show that
a 0.2\% increase is unlikely.  We note that another phenomenon that
could, in principle, result in a small brief wavelength-independent
bump is a triple conjunction of the star, planet and a moon orbiting
the planet \citep[e.g.,][]{sartoretti99, simon10}. A moon orbiting the
planet could also be responsible for a shifted midtransit time
\citep[e.g.,][]{sartoretti99, simon07}. Although the existence of a
moon is an exciting possibility, we caution that it is not likely to
be the case in the current system due to the improbability of a triple
conjunction, and perhaps also the dynamical instability of such a
scenario.

We have presented here the results of an observational multisite
campaign, along with a method to combine the partial ground-based
light curves to obtain a complete transit light curve of a planet with
a long period and long transit duration. Our results in comparison
with the simultaneous {\it Spitzer}\, observations of
\citet{hebrard10} allows an assessment of the quality of multisite campaigns
and develop this method for the future discoveries of additional
systems with similar characteristics.  Such systems will be discovered
by photometric observations of stars known to have planets from RV
surveys, like \six\ itself, and by the spaced-based transiting planet
hunters {\it CoRoT} and {\it Kepler}, capable of continuous
photometric monitoring. The recent discovery of CoRoT-9b
\citep{deeg10}, with a 95 day period and an 8 hour long transit, is an
excellent example. Only two transit events were observed by {\it
  CoRoT}, allowing accurate determination of the \lc\ shape parameters,
while ground-based photometric follow-up observations were needed to
refine the transit ephemeris. Using accurate \lc\ parameters,
determined by space-based observations, to measure a midtransit time
of a ground-based \lc\ is similar to the approach we have taken here.

Ground-based observations are limited by effects of the atmosphere and
short observing windows (the latter leading to the need to combine
data obtained by different instruments), but small to medium
telescopes are relatively easily accessible, compared to space
telescopes. Although forming a collaboration between several
observatories is not a trivial task, those will be motivated by the
increase in transit timing variations with orbital period
\citep[e.g.,][]{holman05, agol05}, thus allowing ground-based
observations to detect the effects of an additional, unseen planet in
the system.

Some of the limitations and difficulties mentioned above can be
minimized by using a network of identical instruments spread around
the globe, allowing for continuous (24/7) monitoring of the target,
especially when the observing windows partially overlap, allowing an 
accurate combination of the partial \lcs. The LCOGT network
\citep[e.g.,][]{brown10, lewis10}, once completed, is meant to be such
a network, in both the Northern and Southern hemispheres. For targets
in Northern (or Southern) positions such as \six, the observational
windows will partially overlap.

\acknowledgments

We thank John Caldwell, Steve Odewahn (McDonald Observatory, TX, USA),
James Otto, Ohad Shemmer (Monroe Observatory, TX, USA) and Anthony
Ayiomamitis (Hellenic Astronomical Union, Greece) for their attempt to
observe \six\ during the \jan\ event, although they were unable to do
so due to bad weather.  
This paper uses observations obtained with facilities of the Las Cumbres Observatory Global Telescope.
The MONET network is funded by the Alfried Krupp von Bohlen und Halbach-Stiftung.
RHO observations were supported by the University of Florida and the
College of Liberal Arts and Sciences. K.D.C.\ is supported by an NSF
Graduate Research Fellowship. This material is based upon work
supported by the National Science Foundation under Grant No. 0707203.
This work is partially based on observations made with the Gran
Telescopio Canarias (GTC), installed in the Spanish Observatorio del
Roque de los Muchachos of the Instituto de Astrof\'isica de Canarias,
on the island of La Palma. The GTC is a joint initiative of Spain
(led by the Instituto de Astrof\'isica de Canarias), the University of
Florida and Mexico, including the Instituto de Astronom\'ia de la
Universidad Nacional Aut\'onoma de M\'exico (IA-UNAM) and Instituto
Nacional de Astrof\'isica, \'Optica y Electr\'onica (INAOE).  KDC, HJD
and EBF gratefully acknowledge the observing staff at the GTC and give
a special thanks to Ren\'e Rutten, Jos\'e Miguel Gonz\'alez, Jordi
Cepa Nogu\'e and Daniel Reverte for helping us plan and conduct the
GTC observations successfully. 
HJD acknowledges support by grant ESP2007-65480-C02-02 of the Spanish
Ministerio de Ciencia e Inovaci\'{o}n.
JNW gratefully acknowledges support from the NASA Origins program through award NNX09AB33G and from the MIT Class of 1942.
This research was partly
supported by the Israel Science Foundation (grant No. 655/07)
and by the United StatesÐIsrael Binational Science Foundation
(BSF) grant No. 2006234.

{\it Facilities:} \facility{FTN (Spectral)}, \facility{GTC (OSIRIS)}


\begin{deluxetable}{cccc}
\tablecaption{\label{tab:lc} HD~80606 light curves presented in this work\tablenotemark{1}.} 
\tablewidth{0pt}
\tablehead{\colhead{HJD} & \colhead{Flux Ratio} & \colhead{Error} &  \colhead{Observatory\tablenotemark{2}}}
\startdata
\hline
\multicolumn{4}{c}{January~2010 transit event}\\ 
\hline
2455210.288327&   1.00114&   0.00116&    WO\\ 
2455210.445392&   1.11691&   0.00024&    GTC\\ 
2455210.446272&   0.99860&   0.00106&    OHP\\ 
2455210.515999&   1.10749&   0.00406&    AO\\ 
2455210.607419&   1.11228&   0.00107&    RHO\\ 
2455210.654098&   1.11259&   0.00110&    FLWO\\ 
2455210.748739&   1.11077&   0.00142&    TMO\\ 
2455210.588711&   1.10630&   0.00116&    WAO-1\\ 
2455210.815045&   1.11449&   0.00134&    WAO-2\\ 
2455209.863305&   0.99977&   0.00037&    FTN\\ 
\hline
\multicolumn{4}{c}{September 2009 transit event}\\ 
\hline
2455098.867697&   1.12197&   0.00227&    RHO\\ 
2455098.950735&   1.11411&   0.00138&    FLWO\\ 
2455098.974510&   1.12694&   0.00101&    MLO\\ 
\enddata
\tablenotetext{1}{ {The table is given in its entirety in the on-line version of the manuscript. A sample is given here, to show its format.}}
\tablenotetext{2}{ {Observatories name abbreviations: 
WO = Wise Observatory, 
GTC = Grand Telescopio Canarias, 
OHP = Observatoire de Haute Provence, 
AO = Allegheny Observatory, 
RHO = Rosemary Hill Observatory, 
FLWO = Fred Lawrence Whipple Observatory, 
TMO = Table Mountain Observatory, 
WAO = Wallace Astrophysical Observatory, 
FTN = Faulkes Telescope North, 
MLO = Mountain Laguna Observatory} }
\end{deluxetable}

\begin{deluxetable}{ccccccccccccc}
\tablecolumns{12}
\tablewidth{40pc}
\tablecaption{\label{tab:obs} Observatories List}
\tablehead{
\colhead{\#} & 
\colhead{Obs.} & 
\colhead{Tel.} & 
\colhead{Filter} & 
\multicolumn{5}{c}{Transit Part\tablenotemark{1}} &
\colhead{$\alpha$} &
\colhead{$\beta$} & 
\colhead{RMS\tablenotemark{2}} &
\colhead{PNR\tablenotemark{3}} \\
\colhead{}  & \colhead{}   & \colhead{[m]}  & \colhead{}   & \multicolumn{5}{c}{OIBEO} & \colhead{} & \colhead{} &\colhead{[\%]}&\colhead{[\% min$^{-1}$]}
}
\startdata
\cutinhead{Jan~2010}
1  & WO   & 1.0    &   $Z$ &                                                      O&I&B&-&-&      1.45&   1.00 & 0.11  & 0.29 \\ 
2  & GTC  & 10.4 &  7680--7780\AA\ ($i'$) &                         -&I&B&E&-&     3.29 &  2.15 & 0.05  & 0.09 \\
3  & OHP  & 1.2   &  Gunn-$r$+ND\tablenotemark{4}  &      -&I&B&-&-&      2.49&   2.27 & 0.13  & 0.25 \\
4  & AO   & 0.41   &  $I$ &                                                        -&-&B&E&O&   1.00&   1.10 & 0.12  &  0.43\\
5  & RHO  & 0.76 &  $i'$ &                                                       -&-&B&E&-&     1.91&   2.59 & 0.09  & 0.15\\
6  & FLWO & 1.2  &  $i'$ &                                                       -&-&B&-&-&      1.11&   1.53 & 0.08   & 0.21\\
7  & TMO  & 1.0    &  $i'$ &                                                       -&-&B&E&O&   1.04&   1.18 & 0.07   & 0.20\\
8  & WAO-1  & 0.36\tablenotemark{5}& $I$ &                       -&-&B&E&O &  1.64&   1.00 & 0.16   & 0.33\\
9  & WAO-2  & 0.36\tablenotemark{5}& $I$ &                       -&-&-&E&O &     1.25&   1.35 & 0.11  & 0.28\\
10 & FTN  & 2.0  & $Z$ &                                                        -&-&-&E&O&      1.24&   1.07 & 0.07   & 0.23\\
\cutinhead{Sep~2009}
1  & RHO  & 0.76 &   $i'$ &                                                      O&-&-&-&-&      1.00&   1.00 & 0.12 & 0.44 \\
2  & FLWO & 1.2  &   $i'$ &                                                      O&I&-&-&-&      1.00&   1.00 & 0.08 & 0.30 \\
3  & MLO  & 1.0  &  $i'$ &                                                         O&I&-&-&-&      1.06&   1.00 & 0.09 & 0.23 \\
\enddata
\tablenotetext{1}{ {OIBEO for Out of transit before ingress, Ingress, flat Bottom, Egress and Out of transit after egress. } }
\tablenotetext{2}{ {rms residual of the 10 minute binned light curve.}}
\tablenotetext{3}{ {Photometric Noise Rate of the unbinned light curve, calculated as rms/$\sqrt{\rm \Gamma}$ where $\Gamma$ is the median number of exposures per minute.}}
\tablenotetext{4}{ {Neutral Density filter.}}
\tablenotetext{5}{ {Two identical telescopes.}}
\end{deluxetable}

\begin{deluxetable}{ll}
\tablecolumns{3}
\tablewidth{0pc}
\tablecaption{\label{tab:tc} Midtransit times.}
\tablehead{
\colhead{Reference} & \colhead{$T_c$ [HJD]}}
\startdata
\hline
\multicolumn{2}{c}{Jan~2010}\\
\hline
This work                   &  2455210.6502(64) \\
H{\'e}brard et al.~2010 &  2455210.6420(10) \\
Winn et al.~2009\tablenotemark{1} &  2455210.6590(51) \\
\hline
\multicolumn{2}{c}{Sep~2009}\\
\hline
This work                   &  2455099.196(26) \\
Winn et al.~2009\tablenotemark{1} &  2455099.2216(50) \\
\enddata
\tablenotetext{1}{ {Predicted $T_c$ based on the ephemeris given in that reference. } }
\end{deluxetable}

\onecolumn

\end{document}